\documentstyle[prl,aps,multicol]{revtex}
\begin{document}
\title{\sc Conductivity of quasiperiodic systems : a numerical study}
\author{\sc S. Roche
\thanks {presently at Department of Applied Physics-University of Tokyo
e-mail : roche@coral.t.u-tokyo.ac.jp}  and D. Mayou}
\address{Laboratoire d'Etudes des Propri\'et\'es Electroniques des Solides,
Centre National de la Recherche Scientifique, B.P.166, F-38042 Grenoble Cedex
9, France.} \date{July 15, 1997}
\maketitle
\widetext
\begin{abstract}
\leftskip 54.8pt
\rightskip 54.8pt
We develop
a new real-space method which allows to evaluate
the {\it Kubo-Greenwood} formula for dc-conductivity of independent electrons in
a static potential.
  We apply it to a numerical study of propagation modes in 3 dimensional
quasiperiodic systems. These modes are strikingly different from those of
periodic ones, with regard to the effect of disorder. In particular for Fermi
energies in pseudo-gaps the conductivity can be stable or can even increase when
disorder increases.
\end{abstract}
\begin{multicols}{2}

Recently, studies on transport in general topologically
disordered media have emphasized the importance to understand how localisation
of states and anomalous quantum transport are related \cite{Simon}. For
example, in
disordered systems, close to a Metal-Insulator transition (at or close to
mobility
edges), localisation aspects of wavefunctions can be described by multifractal
analysis\cite{Hent} and related to critical exponents of
conductivity\cite{Mirlin}.
Another example is the problem of localisation and transport properties in
quasiperiodic systems. This has attracted a growing attention since the
discovery
of quasicrystals in 1984\cite{Schech}. Historically, Kohmoto et
al.\cite{Kho} were
the first to propose that, at least in 1D systems, {\it quasiperiodicity induced
long range correlations giving rise to an intermediate state of localisation},
namely ``critical states" associated to {\it singular continuous} spectra.
Numerical
studies of quantum dynamics of wave packets \cite{Abe} and Landauer resistances
\cite{Kubo} manifested peculiar features such as bounded resistance
$\rho_{N}\leq cste\ N^{\alpha(N,E)}$, for energies E in the spectrum
of the Hamiltonian, with ${\alpha(N,E)}$ encoding the memory of these
``quasiperiodic correlations". More recently, correlations between
quantum dynamics and localisation properties in Fibonacci chains
 have been rigorously analysed by means of renormalisation group
treatments \cite{Piech} and through new numerical methods (Iterated function
systems\cite{Guarneri})

\vspace{5pt}

Experimentally quasicrystalline phases have unique
electronic properties. They are characterized by a low
conductivity, which increases when temperature or disorder increases, and a
proximity to a
Metal-Insulator transition \cite{Poon}. Since real systems allways contain
some defects either
static, due to chemical or structural disorder, or dynamic due to phonons,
it is of
great interest to know how conductivity is affected by disorder in a
quasiperiodic
system.

\vspace{5pt}

In this context, bandstructure calculations of periodic approximants have been
performed. They predict flat bands which are associated to very low Fermi
velocity
. These flat bands are associated to states that have a multifractal character.
\cite{Fuji,Jans}. Estimates of the  conductivity within the Bloch-Boltzmann
theory (and the relaxation time approximation) lead also to small
conductivities.
However the application of the Bloch-Boltzmann theory to these systems has been
criticized \cite{May3,Fuji}. Indeed the propagation of electrons in the perfect
quasiperiodic structure is neither ballistic as in the case of
periodic systems nor diffusive as in the case of disordered systems. One
expects rather
scaling laws of the form $ L(t)= A {t}^{\beta}$ for the extension $L(t)$ of a
wavepacket, where $\beta$ depends on the energy and on the Hamiltonian
parameters. Thus
in a first approximation the diffusivity will be given by $D =
{L(\tau)}^{2}/ 3\tau  = B{\tau}^{2{\beta}-1}$ where $\tau$
 is the finite lifetime induced by
disorder. In the
context of a model this has been confirmed by the mathematical work of
Bellissard and
coworkers \cite{Bell3}. Since states tend to be localised, it has also been
proposed that the mechanism of conductivity can be a hopping mechanism. In that
case, according to the arguments\cite{May3,Fuji}, inelastic or even elastic
scattering
could lead to an increase of the conductivity.

\vspace{5pt}

In order to go beyond the Bloch-Boltzmann description, and be
able to test the various schemes that have been proposed, there is a natural
starting point given by the linear response theory. This treatment does not
make any assumption on the transport mechanism. In this Letter we present
the first
study, to our knowledge, of Kubo-Greenwood conductivity for independent
electrons in
static quasiperiodic potential at $T=0K$. Using a new real-space method, it
will be shown that the electronic conduction in quasiperiodic systems differs
strinkingly from the prediction of a Bloch-Boltzmann approach. In
particular if $V_{dis}$
measures the amplitude of the static disordered potential we will show that the
conductivity does not vary like $\sigma = \sigma_{0}/ {V_{dis}}^{2}$ (where
$\sigma_{0}$ is independent of disorder) in the weak
scattering limit. Furthermore the variation of conductivity with disorder is
rather complex, and in particular it depends strongly on the position of
the Fermienergy with respect to pseudo-gaps. \vspace{5pt}

\vspace{1pt}

\hspace{\parindent}For independent electrons in a given static potential at
$T=0K$, the starting point of the method will be the following form of
Kubo-Greenwood  ($E$ is the Fermi energy) : {\small
$$\sigma_{DC}(E)=\frac{2\hbar {e}^{2}\pi}{\Omega}\hbox{Tr}\bigl[
 \hat{V}_{x}\ \delta(E-{\cal H})\ \hat{V}_{x}\ \delta(E-{\cal H})\bigr]$$}

$\hat{V}_{x}$ is component of the velocity operator along direction x.
$\delta(E-{\cal
H})$ is the projector on eigenstates {\cal H} of energy E. $\Omega$ is the
volume
of the system. The factor 2 comes from the spin degeneracy. In this work, the
evaluation of the Kubo-Greenwood formula is made through a new  real space
method.
In a previous work \cite{mayou} we developped a method for calculating dc and ac
conductivity using the formalism of orthogonal polynomials. Here, based on
the same
formalism, we developp a method which gives access only to dc-conductivity. In
counterpart this allows an important gain in precision and numerical
stability. For
convenience, after some simple algebra one can rewrite
{\small$\sigma_{DC}(E)$}  as :
$${\small\sigma_{DC}(E)}= \lim_{t\to\infty}F(E,t)$$

{\small$$F(E,t) = \frac{2\hbar {e}^{2}\pi}{\Omega}\hbox{Tr} \bigl[
\frac{1}{t}
 \bigl(\hat{X}(t)-\hat{X}\bigr) \ \delta(E-{\cal H})\
\bigl(\hat{X}(t)-\hat{X}\bigr) \bigr]$$} \noindent
Here ${\small\hat{X}(t)}= {e}^{i{\cal
H}t/\hbar}{\small\hat{X}}{e}^{-i{\cal H}t/\hbar}$ where ${\small\hat{X}}$
is the
component along direction x of the position operator. At this stage, we
define for
each orbital $\small\mid\ j>$ :

  $${\small\mid
\Phi_{j}(t)>}= \hat{X} e^{-i{\cal H}t/\hbar}{\small\mid\ j>}$$
{\small $$\mid
\widetilde\Phi_{j}(t)>=\frac{\mid \Phi_{j}(t)>} {\parallel\!  \ \mid
\Phi_{j}(t)>\ \!\parallel}$$}
Here $\small\mid\widetilde\Phi_{j}(t)>$ is a normalised state. The
calculation of conducivity can be reduced to :

{\small$$\frac{2\hbar {e}^{2}\pi}{\Omega} \sum_{j} {\cal D}_{j}(t)\times
<\widetilde{\Phi}_{j}(t)
 \mid \delta(E-{\cal H}) \mid \widetilde\Phi_{j}(t)>$$}

For each initial state at $\mid j>=\mid \Psi
(t=0)>$, we can take the origin at j. This means that  $<j\mid\hat{X}\mid
j>=0$. Then ${\cal D}_{j}(t)$ is given by

{\small$${\cal D}_{j}(t)= \frac{<\Psi_{j}(t)\mid \hat{X}^{2} \mid \Psi_{j}(t)>}
{t}$$}.

\vspace{5pt}

The conductivity can thus be calculated from ${\cal
D}_{j}(t)$ and the spectral quantity related to
{\small$\mid\widetilde{\Phi}_{j}(t)>$}. The
spectral quantity is calculated by the recursion method  \cite {Hayd}. The
original part of
our method is the evaluation of $\mid \Psi_{j}(t)>$. Our method, to solve
the Schr\"{o}dinger
equation, avoids to use Runge-Kutta resolution algorithms, or numerical
procedures of
diagonalisation. It is based on the development of $\delta(E-{\cal H})$ already
used in \cite{mayou}. which implies also a development of the unitary operator
${e}^{-i{\cal H}t/\hbar}$ on a basis of orthogonal polynomials. Hereafter,
we will
take Chebyshev polynomials of the first kind associated to the weigth
$\rho(E)=1/(\pi\sqrt{\strut 4b_{\infty}^{2}-(E-a_{\infty})^{2}})$  and
defined via the
recursive relations :

\begin{eqnarray}
Q_{0}(E)&=& \ 1 \ \ \ Q_{1}(E)=\ \frac{E-a_{\infty}}{2b_{\infty}}\nonumber\\
Q_{n+1}(E)&=&\biggl(\frac{E-a_{\infty}}{b_{\infty}}\biggr)
Q_{n}(E)-Q_{n-1}(E)\nonumber
\end{eqnarray}

\noindent
where $a_{\infty}$ and $b_{\infty}$ are chosen as band parameters in
accordance with the ones of the true density of states (via a first recursion
process). Then the vector under study reads :

\begin{eqnarray}
{e}^{-\frac{i{\cal H}t}{\hbar}}\mid j>&=&\sum_{n} h_{n}  \bigl( \int dE
\rho(E)Q_{n}(E) {e}^{ -\frac{iEt}{\hbar}} \bigr ) Q_{n}({\cal H})\mid
j>\nonumber \\ &=&\sum_{n} h_{n} {i}^{n}J_{n}\bigl( \frac{-2b_{\infty}t}{\hbar}
\bigr)\  {e}^{\frac{-ia_{\infty}t}{\hbar}}Q_{n}({\cal H})\mid j> \nonumber
\end{eqnarray}
		
$h_{0}=1$ and for other $n$ $h_{n} = 1/2 $. As usual with orthogonal
polynomials the
{\small $Q_{n}({\cal H})\mid j>$} are evaluated via the
 recurrence property \cite{Hayd}. Amplitudes of $ {e}^{-\frac{i{\cal
H}t}{\hbar}}\mid j>$ on this basis have a rather simple form connected to Bessel
functions. The consequent interest is to get nice asymptotic behaviors for these
coefficients which converge very quickly as  $n\to\infty$, given that
$\lim_{n\to\infty} J_{n}(z)\sim \frac{1}{\sqrt{2\pi n}} \bigl(\ \frac{ez}{2n}\
\bigr)^{n}$. It is this development that makes the calculation quick and
precise.

\vspace{5pt}

For each $j$ the calculation is performed on a
cube, centered on the site $j$, which length is $100$ sites (the
cube contains about ${10}^{6}$ sites). Also, for pratical calculations, the
sum is performed
over $\sim 100$ sites of origin. We find that this is sufficient for our
purpose. We have
performed numerous test (more details can be found elsewhere \cite{Roche})
to ensure that the
convergence is achieved. We estimate that the conductivity is calculated
with an energy
resolution of a few percent of the bandwidth which is enough for our
purpose. As for the
calculation of conductivity the total computing time is of the order of
$100$ hours on a HP$735$
for each conductivity curve $ \sigma_{DC}(E)$ (see below).

\vspace{5pt}

\hspace{\parindent}We performed calculations of quantum
 diffusion and  Kubo-Greenwood conductivity at $T=0K$.
We consider a s-band tight-binding model on a simple cubic
lattice with nearest-neighbor hopping which has already been studied by several
authors \cite{Sir4,Zhong}. The hopping integral is  the energy unit ($t=1$)
and the on-site energies are given by
$\varepsilon_{j}=\varepsilon_{x_{j}}+\varepsilon_{
y_{j}}+\varepsilon_{ z_{j}}+ \varepsilon_{dis}$ with
$\varepsilon_{dis} = \ \hbox{random number} \
\in \bigl[-\frac{V_{dis}}{2},+\frac{V_{dis}}{2}\bigr]$ and
$\varepsilon_{j_{\alpha}}=\pm V_{qp}$
constraints to quasiperiodic correlations (Fibonacci sequence). This model
allows a
direct comparison between a quasiperiodic system and a periodic one since for
$V_{qp}=0$ one recovers the classical Anderson model with diagonal disorder. We
note also that for $V_{dis}=0$ the Hamiltonian is separable. An eigenstate
 $\Psi(x,y,z)$
can be written as the product of eigenstates of the chains along each
direction $\Psi(x,y,z)=\Psi_{1}(x)\times \Psi_{2}(y)\times
\Psi_{3}(z)$ the energy being the sum of the three energies $ E = E_{1}+ E_{2}+
E_{3}$.  This means also that for a state that is initially localised on a
site j
one has $\Psi_{j}(x,y,z,t)=\Psi_{1}(x,t)\times \Psi_{2}(y,t)\times
\Psi_{3}(z,t)$ with $ \Psi_{1}(x,t=0) = \delta(x - x_{j}) $ and similarly
for y and z. Obviously the Hamiltonian is no more separable when $V_{dis}$
is non zero.
\vspace{5pt}

We study the quantum diffusion through $D_{j}(t)$. For $V_{dis} = 0$ the above
relation shows that $D_{j}(t)$ is the same as for a one dimensional
model\cite{Piech}. In this study we focus on the effect of disorder. Our
results (see figure $1$) , clearly show
that conduction modes undergoes a transition from non-ballistic to
diffusive regime
 (i.e. $D_{j}(t)$ is independent of t at large t). When the disorder
increases the  transition to the diffusive regime occurs at shorter times
and the
asymptotic value of $D_{j}(t)$ tends to decrease. We note also that the
fluctuations of $D_{j}(t)$ are less important when the disorder increases.

\vspace{10pt}

We studied also the conductivity $\sigma_{DC}(E)$ and its
variation with the strength of disorder. In a metal one expects a law of
the form
$\sigma = \sigma_{0}/ {V_{dis}}^{2}$ ($\sigma_{0}$ is independent of
disorder). Indeed
this is what we find for $V_{qp} = 0$. For sufficiently small values of
disorder our
numerical results are in good agreement with the prediction of the
Bloch-Boltzmann
approximation \cite{Roche}. However, the enhancement of disorder in a
quasiperiodic
system leads to a different law. For instance in the figure $2$ the conductivity
decreases when the disorder increases but does not follow the law $\sigma =
\sigma_{0}/
{V_{dis}}^{2}$ (note that the density of states is nearly independent of
disorder in
this parameter range). Instead we get approximately for each energy E a law
of the form
$\sigma = \sigma_{0}/ {V_{dis}}^{\alpha}$. Our results do  not strictly
follow a power
law, probably because such laws apply only to the limit of infintely small
disorder, and
also due to the finite accuracy of our method. Depending on  the energy E
we find
$\alpha = 0.4 - 0.8 $ in our calculation.

\vspace{5pt}

In figure $3$, we show the conductivity for a quasiperiodic modulation
$V_{qp}=1.1$. A striking  result is that there are particular zones,
identified by
pseudo-gaps which seem quite insensitive to a  tremendous increase of
disorder (a factor of 16 for ${V_{dis}}^{2}$). The conductivity varies
monotonically for intermediate values of disorder not shown here. The inset
in figure
$3$ shows the average density of states. The density of states
increases slightly with disorder in the pseudo-gaps. This increase
compensates for the
decrease of diffusivity leading to a nearly constant conductivity. In
regions of high
density of states, which correspond also to higher conductivity, both
density and
diffusivity decrease.

\vspace{5pt}

In figure $4$ we show {\small$\sigma_{DC}(E)$} for $V_{qp}=2.5$ and different
values of $V_{dis}$ (again the conductivity varies
monotonically for intermediate values of disorder not shown here). The
variation of
conductivity with energy is rather complex. We note however that the
conductivity increases with $V_{dis}$ for some energies that correspond to
pseudo-gaps. In contrast to the previous cases there is an important change
of the
density of states. It increases where the conductivity increases. It means
that the
electronic structure is deeply modified by the disorder. Thus we prove that the
localization induced by a quasiperiodic potential can be destroyed by a
disordered
potential. \vspace{5pt}

A direct comparison with experimental results on
quasicrystals is difficult, since our model does not provide a realistic
description of the electronic structure and of the local atomic order.
However, quite
remarquably, when the Fermi energy lies in a pseudo-gap, which is the case
experimentally, the behavior of the conductivity is reminiscent of the
experimental observation that the conductivity increases with disorder or with
temperature \cite{Poon}.

\vspace{5pt}

\hspace{\parindent} To conclude, the contribution of this paper is twofold. We
have developped a new real-space method which allows to study electronic
propagation and to calculate the conductivity of quasiperiodic systems
(through the
{\it Kubo-Greenwood} formula). This opens new ways of investigating the
transport
properties of these systems, but of course the method is not restricted to
quasiperiodic potentials. Furthermore our first numerical study of
conductivity demonstrates the complexity of electronic transport in
quasiperiodic
systems with disorder. In the context of anomalous localisation, it shows how a
quasiperiodic potential can affect quantum transport in a specific way,
when compared
to a periodic or a disordered potential. For sufficiently small disorder and
quasiperiodic potential the reduction of conductivity is in qualitative
agreement with
the prediction of scaling laws. However a particularly interesting result is the
inhomogeneous variation of conductivity with Fermi energy. In zones of
pseudo-gaps the
conductivity can be stable upon enhancement of disorder or can even
increase with
disorder.

\vspace{10pt}
{\bf ACKNOWLEDGEMENTS}

\vspace{5pt}
This work was partially supported by a NATO grant CRG 941028 which is gratefully
acknowledged.

\vspace{10pt}
{\bf FIGURE CAPTIONS}

\vspace{10pt}
{\bf Figure 1} : ${\cal D}_{j}(t)$ represented for several initial sites j and
$V_{qp}=0.9$. Time is in units of $ 2\hbar/W$ where $W$ is the bandwidth.
Lengthes are in unit of the nearest-neighbors distance. (a)$V_{dis}=2$ (b)
$V_{dis}=2\sqrt{2}$. The thick line is an aid to visualize one of the curves.

\vspace{10pt}
{\bf Figure 2} : Conductivity {\small$\sigma_{DC}(E)$} for $V_{qp}=0.7$ (average
value over sites "j" of $\lim_{t\to\infty} D_{j}(t)N_{j}(E)$ where
$N_{j}(E)$ is the
spectral weight on the state  $\small\mid\widetilde\Phi_{j}(t)>$). The
energy unit is
the hopping integral t. Inset : average density of states. (a) $V_{dis}=2$.
(b) $V_{dis}=2\sqrt{2}$

\vspace{10pt}
{\bf Figure 3} : Conductivity {\small$\sigma_{DC}(E)$} for $V_{qp}=1.1$  as a
function of the Fermi energy $E$ for different values of the disorder parameter.
Inset : average density of states. (a) $V_{dis}=1/\sqrt{2}$ (b)
$V_{dis}=2\sqrt{2}$.

\vspace{10pt}
{\bf Figure 4} : Conductivity {\small$\sigma_{DC}(E)$} for $V_{qp}=2.5$  as a
function of the Fermi energy $E$ for different values of the disorder parameter.
Inset : average density of states. (a)$V_{dis}=2$ (b) $V_{dis}=4\sqrt{2}$.

\end{multicols}
\end{document}